\documentclass[floatfix,onecolumn,showpacs,amsmath,amssymb]{revtex4}

\usepackage{epsf}
\usepackage{graphicx}  
\usepackage{dcolumn}   
\usepackage{bm}        
\usepackage{color}

\newcommand{\be}{\begin{equation}}
\newcommand{\en}{\end{equation}}
 \newcommand{\bea}{\begin{eqnarray}}
 \newcommand{\ena}{\end{eqnarray}}
  \newcommand{\sch}{Schwarzschild}

\begin{document}

\title{Spherical gravitational waves and quasi-spherical waves scattered from black string in massive gravity}
\author{ Hongsheng Zhang$^{1}$}\email{sps_zhanghs@ujn.edu.cn}
\author{Yang Huang $^1$}\email{sps_huangy@ujn.edu.cn}
\affiliation{$^1$ School of Physics and Technology, University of Jinan, 336 West Road of Nan Xinzhuang, Jinan, Shandong 250022, China}

\date{ \today}

\begin{abstract}
   Spherical gravitational wave is strictly forbidden in vacuum space in frame of general relativity by the Birkhoff theorem. We prove that spherical gravitational waves do exist
   in non-linear massive gravity, and find the exact solution with a special singular reference metric. Further more, we find exact gravitational wave solution with a singular string by meticulous studies of familiar equation, in which the horizon becomes non-compact. We analyze the properties of the congruence of graviton rays of these wave solution. We clarify subtle points of dispersion relation, velocity and mass of graviton in massive gravity with linear perturbations. We find that the graviton ray can be null in massive gravity by considering full back reaction of the massive gravitational waves to the metric. We demonstrate that massive gravity has deep and fundamental discrepancy from general relativity, for whatever a tiny mass of the graviton.

\end{abstract}


\preprint{arXiv: }
 \maketitle

 \section{Introduction}
 Neither equivalence nor general covariance principle is the perfect principle of general relativity. All metric theories satisfy weak equivalence principle and Einstein equivalence principle \cite{will}. General covariance
 principle has no essential constraint on physical laws \cite{kre}. In fact, Newtonian gravity also satisfies this principle without intrinsical physical modifications. The principle which uniquely leads to general relativity is the following statement. General relativity is the non-linear theory of massless spin-2 particle. One can demonstrate that this statement implies general relativity \cite{win}. In this view, a most natural extension of general relativity is to consider the theory of massive spin-2 particle. To find a ghost-free non-linear massive gravity is not an easy task. A satisfactory non-linear ghost-free massive gravity is recently suggested by de Rham, Gabadadze and Tolley (dRGT) \cite{drgt}, and the property of ghost-free is examined by successive expansion to the 4th order. Then a complete demonstration of the property of ghost-free, which does not depends on linear expansions of dRGT massive gravity, is proposed in \cite{hassan}.

 In vacuum space, the equation of motion of massless spin-2 particle, or linear general relativity (GR), reads,
 \be
 \Box h_{ab}=0,~~\partial^a h_{ab}=0,~~h_a^a=0,
 \label{linearh}
 \en
 where $h$ is the fluctuation around a background metric, taken as Minkowskian metric in the simplest case.
 A ghost-free linear theory of massive gravity is proposed in \cite{FP}, in which the equation of motion of metric fluctuation becomes,
 \be
  \Box h_{ab}-m^2h_{ab}=0,~~\partial^a h_{ab}=0,~~h_a^a=0.
 \label{massh}
 \en
 Clearly, a spherical wave
 \be
 h_{ab}(t,r)=e_{ab}\left(c_1\frac{f_1(t-r)}{r}+c_2\frac{f_2(t+r)}{r}\right),
 \en
 solves the linear equation of motion in GR (\ref{linearh}), where $e_{ab}$ is a constant tensor, $f_1, f_2$ are arbitrary functions of the arguments, and $r$ is the standard areal coordinate. However, due to
 Birkhoff's theorem, spherical wave is strictly forbidden in full theory of GR. Thus the property of a full non-linear theory may be fundamentally deviated from the corresponding linear approximation. Or linear approximation
  may lose essential information of the full theory.
 For the linear massive gravity, one has spherical wave solution,
 \be
 h_{ab}(t,r)=e_{ab}\left(c_1\frac{e^{-i(\omega t-kr)}}{r}+c_2\frac{e^{-i(\omega t+kr)}}{r}\right),
 \en
 where $\omega^2=k^2+m^2$. From the lesson of the relation between GR and linear GR, we have no idea of the spherical wave in full non-linear massive gravity even though it exists in linear massive gravity.

 Modified gravity corresponds to GR with some special sources, and thus the Birkhoff theorem fails. A spherical wave does not be ruled out in principle. We shall demonstrate that it really exists in massive gravity, and present the exact solution. Exact solution plays a significant role in gravity because of its non-linearity. Exact solution dramatizes some inherent properties of the theory, while the linear approximation cannot fulfill this task under several situations. In 1925, the first exact solution for gravitational wave in GR is found in \cite{brink}. It is a solution of plane-fronted gravitational wave with parallel ray (pp-wave), which can describe plane gravitational wave with arbitrary amplitude and arbitrary wave form.  Spherical wave, as a most useful model for other interactions and in linearized general relativity, is absent in general relativity for vacuum space. Topological spherical waves in $f(R)$-gravity are found in \cite{self1}. We find that topological spherical waves appear for an infinitesimal modification of general relativity. This property seems remarkable since it is ``quantized". The theory permits topological spherical wave for any tiny deviation from GR. The answer of existence of such a wave is ``yes" or ``no" rather than to constrain a parameter to an interval. A spherical wave is more realistic in the universe (however see \cite{topo}). We shall show that spherical wave exists when the graviton is endowed with mass.  From aspect of observations, one distinguishes general relativity from modified gravity even for one certain test by invoking such a property.

  \section{spherical gravitational waves in dRGT massive gravity}
  We start from the dRGT action,
  \be
\label{action}
S =\int d^{4}x \sqrt{-g} \left[ \frac{1}{16\pi G}\left(R +V(g,f)\right)+{\cal L}_m\right],
\en
     where $R$ denotes the Ricci scalar, ${\cal L}_m$ labels the Lagrangian of the matter fields, and $V(g,f)$ is the potential of graviton. $V(g,f)$ is a function of the physical metric $g$ and reference metric $f$, which is the critical improvement of dRGT massive gravity. The reference metric is independent of the physical metric. The potential $V$ reads,
      \be
       V(g,f)=m^2 \sum^4_i c_i {\cal U}_i (g,f),
        \label{poten1}
        \en
          where,
          \bea
\label{poten}
&& {\cal U}_1= [{\cal K}], \nonumber \\
&& {\cal U}_2=  [{\cal K}]^2 -[{\cal K}^2], \nonumber \\
&& {\cal U}_3= [{\cal K}]^3 - 3[{\cal K}][{\cal K}^2]+ 2[{\cal K}^3], \nonumber \\
&& {\cal U}_4= [{\cal K}]^4- 6[{\cal K}^2][{\cal K}]^2 + 8[{\cal K}^3][{\cal K}]+3[{\cal K}^2]^2 -6[{\cal K}^4].
\ena
       The new tensor $\cal K$ is defined as,
       \be
       {\cal K}^{\mu}_{\nu}= \left(\sqrt {g_M^{-1}f_M}\right) ^{\mu}_\nu.
       \label{root}
       \en
         Here $g_M^{-1}$ denotes $g^{\alpha\beta}$ in matrix form, and $f_M$ denotes $f_{\alpha\beta}$ in matrix form. $[{\cal K}]$ labels the trace of ${\cal K}$. $c_i$ are four constants. The rigorous study in \cite{hassan2} shows that the quartic term ${\cal U}_4$ can be reduced to ${\cal U}_3$, ${\cal U}_2$ and ${\cal U}_1$ in field equation. Thus we can always set $c_4=0$ without loss of generality.    A special note is that the root operation of a matrix is of subtle.  For example the most simple matrix diag(1,1) has at least the following different square roots:  diag(1,1), diag(1,-1), diag(-1,1), and diag(-1,-1), and the Pauli matrices. This subtlety is discussed in detail in \cite{hassan3}.  The equation of motion of the physical metric $g$ corresponding to the action (\ref{action}) reads,
   \begin{eqnarray}
R_{\mu\nu}-\frac{1}{2}Rg_{\mu\nu}+m^2 \chi_{\mu\nu}&=&8\pi G T_{\mu \nu },~~
\label{field}
\end{eqnarray}
where $T_{\mu\nu}$ is the matter term corresponding to ${\cal L}_m$, $\chi _{\mu\nu}$ is the potential term of the gravity field,
\begin{eqnarray}
&& \chi_{\mu\nu}=-\frac{c_1}{2}({\cal U}_1g_{\mu\nu}-{\cal K}_{\mu\nu})-\frac{c_2}{2}({\cal U}_2g_{\mu\nu}-2{\cal U}_1{\cal K}_{\mu\nu}+2{\cal K}^2_{\mu\nu})
-\frac{c_3}{2}({\cal U}_3g_{\mu\nu}-3{\cal U}_2{\cal K}_{\mu\nu}\nonumber \\
&&~~~~~~~~~ +6{\cal U}_1{\cal K}^2_{\mu\nu}-6{\cal K}^3_{\mu\nu})
-\frac{c_4}{2}({\cal U}_4g_{\mu\nu}-4{\cal U}_3{\cal K}_{\mu\nu}+12{\cal U}_2{\cal K}^2_{\mu\nu}-24{\cal U}_1{\cal K}^3_{\mu\nu}+24{\cal K}^4_{\mu\nu}).
\end{eqnarray}

 We write the metric ansatz of a gravitational wave as follows,
 \be
 g=\eta+S_w,
 \label{total}
 \en
 where $\eta$ denotes the Minkowski metric,
 \be
 \eta=-du^2-2dudr+r^2(d\theta^2+\sin^2\theta d\phi^2),
 \label{eta}
 \en
 and $S_w$ denotes a term of gravitational wave. Inspired by the Brinkmann's pp-wave, we take,
 \be
  S_w=\frac{2U(u)P(\theta,\phi)}{r}du^2.
  \en
  $U(u)$ describes an outgoing wave, and $P(\theta,\phi)$ displays its polarization. It is clear that when $U,~P$ are constants, this metric reduces to a \sch~ one.

   Now we discuss the reference metric. Up to now we have no physical
  principle to determine the reference metric. Sometimes it is taken as a Minkowski one. However, no sound physical reason prohibits other forms of reference metric.   In fact the Minkowski reference metric is not even the simplest choice. A singular reference metric is also  permitted, and sometimes is required \cite{vegh}. A black hole solution in massive gravity with singular reference metric is found in \cite{cais}, and is widely applied in \cite{several}. The  stability of massive gravity with general singular reference metrics thoroughly explored in \cite{self2}. Singular reference metric plays a critical role in some intricate problem. For example, singular reference metrics overcome the awful static universe problem in massive cosmology,  and lead to a sound massive cosmology \cite{mascos}. Here we adopt a singular reference metric as follows,
  \be
  f=\frac{U^2}{r^6}\left(\left(\frac{\partial P}{\partial \theta}\right)^2+\frac{{1}}{\sin^2\theta}\left(\frac{\partial P}{\partial \phi}\right)^2\right)du^2.
  \label{refm}
  \en
  Thus we obtain $\cal K$ by using (\ref{root}),
  \be
  {\cal K}=\frac{2}{r^2}\left({P}dU-UdP\right)du=\frac{2U}{r^2}\left(P\frac{\dot{U}}{U}du^2-\frac{\partial{P}}{\partial \theta}dud\theta-\frac{\partial {P}}{\partial \phi}dud\phi\right).
  \en
  It is clear that Rank$(f)=1$, and Rank$({\cal K})=3$.
  One confirms that,
  \be
  {\cal K}^a_b{\cal K}^b_c=g^{ab}f_{bc}.
  \en

  For vacuum space, one finds that the field equation (\ref{field}) requires,
  \be
  U(u)\left[\frac{\partial^2P}{\partial \phi^2}+\sin\theta\left(\cos \theta \frac{\partial P}{\partial \theta} +\sin\theta \frac{\partial^2P}{\partial \theta^2}\right)\right]=0,
  \label{holo}
  \en
   in which we set $c_1=2/m^2$, and $c_2=0$.
  $U=0$ simply implies a Minkowski space. The terms in the square bracket implies a real non-trivial gravitational wave. It is easy to see that $P$ is  spherical harmonic function of $\theta,~\phi$ with $l=0$. Therefore the unique solution of $P$ is a constant, and $U$ is an arbitrary function of $u$. Apparently, it is the same as a Vaidya metric. The fundamental difference roots in the source. Here, in massive gravity it is a vacuum solution. Contrarily, in GR it is sourced by null dust. And $U$ is treated as the Misner-Sharp mass of a shining black hole. The variation of Misner-Sharp mass of the black hole is exactly balanced by energy carried by the null dust from the black hole to infinity. In massive gravity $U$ also denotes the Minser-Sharp mass \cite{humass}, while there is no matter outside the black hole to balance the variation of the mass of the hole. The only possible explanation is that there are energies carried by gravitational waves from the hole to infinity for a decreasing $U$. For increasing $U$ we need to write the wave solution in advanced Eddington coordinates system. The physical interpretations mimic the case of decreasing $U$.  We thus complete a spherical gravitational wave in massive gravity for vacuum space.

   \section{quasi-spherical waves scattered from black strings}
   Before exploring the propagation properties of this wave, we reveal hidden solutions of (\ref{holo}). The view of  \textit{l=0 spherical harmonics } of (\ref{holo}) comes from ordinary physics. In gravity theories, the situations become different. Under very general conditions, singularities in gravity theory are inevitable \cite{penhaw}. We should discuss singularities seriously in gravity theories. Now we make a cautious analysis of (\ref{holo}). Some special conditions have been imposed for spherical harmonics, which requires that $P$ is periodic with respect to $\phi$ and regular for $\theta\in [0,\pi]$. It is a natural condition that $P$ is periodic of $\phi$. It seems no space to modify this condition. On the other hand, the regularity of $P$ deserves reconsidering. To loose this condition leads to divergent points for special $\theta$. This is a familiar result in theories of partial differential equations.  Here, one will see that such a loosen carries incisive and profound physical results.  To obtain a more general analytical form of the equation (\ref{holo}), we introduce,
  \be
  v=-\ln (\csc \theta+\cot \theta).
  \en
  With this new definition, the equation (\ref{holo}) becomes,
  \be
 U(u)( \frac{\partial^2 P}{\partial \phi^2}+\frac{\partial^2 P}{\partial v^2})=0.
  \en
   It is a two dimensional harmonic equation of $P$, and $U$ is an arbitrary function of $u$. The general solution of $P(v,\phi)$ is real part or imaginary part of a holomorphic function. For example,
  a simple holomorphic function $e^{-z}$ leads to,
  \be
  P_1=(\csc \theta+\cot \theta)\cos\phi,
  \label{P1}
  \en
  and,
  \be
  P_2=(\csc \theta+\cot \theta)\sin\phi.
  \label{P2}
  \en
 One confirms that $P_1$ and $P_2$ preserve the periodic property with respect to $\phi$. And they respectively represent the two polarizations of this spherical wave \cite{sacwu}. A singular string appears at $\theta=0,~\pi$. It is a true singularity because,
 \be
 R^{abcd}R_{abcd}=\frac{48U^2P^2}{r^6}.
 \en
 Thus either singular point of $U,~P$ or $1/r$ is singular point of the spacetime. The metric ansatz (\ref{total}) with solution (\ref{P1}) or (\ref{P2}) is a gravitational wave with a singular string, in which the horizon looks like a bottle with bi-port at both ends. This spacetime permits no Killing vector. We just name it quasi-spherical gravitational wave, based on the property that it reduces to a spherical wave when $P$ is a constant.  The physical interpretation is that the energies required for transverse polarization cause the spacetime to collapse to form a singular string. As a comparison, the spherical wave without transverse polarization, i.e.,  $P=$constant,  only has a singular point at $r=0$.

  A problem for (\ref{P1}) and (\ref{P2}) is that naked singularities appear at $\phi=\frac{\pi}{2}, ~\frac{3\pi}{2}$, or $\phi=0,~\pi$. This is not difficult to overcome. We set the following holomorphic function,
  \be
  H=A+e^{-z},
  \en
  where $A=a_1+ia_2$ is a constant complex number with $a_1a_2\neq 0$, and $z=v+i\phi$. Then,
  \be
  P_1=a_1+(\csc \theta+\cot \theta)\cos\phi,~~P_2=a_2+(\csc \theta+\cot \theta)\sin\phi.
  \en
  With these new $P_1,~P_2$, naked singularity disappears.


  We study the propagation properties of this quasi-spherical gravitational wave. The wave vectors form a congruence of graviton ray, in analogy to light ray. The out going wave vector reads,
  \be
  K_a=-(du)_a.
  \en
  To discuss the properties of the null congruence, we use tetrad formulism for this metric,
  \be
  \tau_1=-[g_{00}du+2g_{01}dr]\frac{1}{2},~~\tau_2=du,~~\tau_3=rd\theta,~~\tau_4=r\sin \theta d\phi.
  \en
  Under this tetrad formalism, the metric becomes,
  \be
  g=-\tau_1\tau_2-\tau_2\tau_1+\tau_3^2+\tau_4^2.
  \label{tetrad1}
  \en
  This graviton congruence of quasi-spherical waves is a null geodesic congruence, for
  \be
  K^aK_a=0,~~K^a\nabla_a K_b=0.
  \en
  Compared to a pp-wave, $K$ is no longer to be a Killing vector,
  \be
  \nabla K=-\frac{U(u)P}{r^2}\tau_2^2+\frac{1}{r} (\tau_3^2+\tau_4^2).
  \label{naK}
  \en
  The deformation of the congruence is embodied in its expansion, shear, and twist on sectional 2-surface $\tau_3-\tau_4$. In the following equations a hat denotes the projection of a tensor on the section $\tau_3-\tau_4$. From (\ref{naK}), we derive the expansion, shear, and twist,
  \be
  \hat{\theta}=\hat{g}_{ab}(\nabla^a K^b){\hat{}}=\frac{2}{r},
  \en
  \be
  \hat{\sigma}_{ab}=(\nabla_{(a} K_{b)}){\hat{}}-\frac{1}{2}\hat{\theta}\hat{g}_{ab}=0,
  \en
  and,
  \be
  \hat{\omega}_{ab}=(\nabla_{[a} K{_b]}){\hat{}}=0,
  \en
    respectively. Therefore, the congruence of this quasi-spherical waves is shear-free, twist-free, but expanding. This result is expected from physical instinct for congruence of radial gravitons. About the ratio of variation of $\hat{\theta}$ with respect to an affine parameter, the Raychaudhuri equation presents,
  \be
  K^a\nabla_a \hat{\theta}=-\frac{1}{2}\hat{\theta}^2-\hat{\sigma}_{ab}\hat{\sigma}^{ab}+ \hat{\omega_{ab}}\hat{\omega^{ab}}-R_{ab}K^aK^b=-\frac{2}{r^2}.
  \en
  The congruence has been affine parameterized since $K^a\nabla_a K_b=0$. $\hat{\sigma}_{ab}$ and $\hat{\omega}_{ab}$ are identically equal to zero all over the congruence.
  It is clear that both expansion and its ratio of variation is independent on wave form $U(u)$ and polarization $P(\theta,\phi)$.

  Up to now all of our discussions of this quasi-spherical wave are exact. All the back reactions of the wave to the metric are taken into account. One may be curious why the wave vector is null and where the effects of the massive graviton are. In the followings we clarify some puzzles in the interpretations of the dispersion relation, mass, and velocity of gravitons in massive gravity when one considers linear approximations.

  To explore the effective graviton mass in this study, we make a linear approximation of the field equation (\ref{field}) for the specific singular reference metric (\ref{refm}) under the conditions $U\ll r$ and $P\ll r$. Under these conditions, the field equation becomes,
  \be
  \Box h_{ab}+\frac{m^2c_1}{2}{\cal K}_{ab}=0.
  \en
  Thus it is clear that a longitude massive graviton with effective mass,
  \be
  m_g^2=-\frac{c_1m^2\dot{U}P}{r^2},
  \en
  appears. This equation is similar to the result in \cite{imm}.

  Equation (\ref{total}) decomposes the whole metric into a Minkowski background $\eta$ and a wave $S_w$. Clearly, $S_w$ is really a null wave. We emphasize that this is the result by considering full back reactions of the wave to the Minkowski metric. In the scenario of linear gravity theory,  a wave propagating on the background of a Minkowski spacetime. The critical point is that $u$ in (\ref{eta}) is, in fact, the retarded Eddington coordinate, and $r$ is the areal coordinate rather than tortoise coordinate. To get back into the scenario of linear gravity, and thus put the wave on a Minkowski spacetime, we need to recover the spherical coordinate $r$ in $U(u)$ to measure the dispersion relation of the radial gravitons. Once the physical picture is cleared, the calculation is easy and straightforward,
  \be
  S_w=\frac{2U(t-r_*(r))P}{r}du^2,
  \en
  where
  \be
  r_*=r+2UP\ln \frac{r-2UP}{2UP}.
  \label{rstar}
  \en
  (\ref{rstar}) is a recursive equation of $r_*$ since $U$ is a function of $r_*$. For linear approximation and a slowly varying $U$, we assume
  \be
  UP\ll r.
  \en
  Thus, we have,
  \be
  \frac{dr_*}{dr}=1+\frac{2UP}{r}.
  \en
  So, the apparent dispersion relation measured by the areal coordinate $r$ reads,
  \be
  \frac{1}{v_g}=\frac{k}{\omega}=1+\frac{2UP}{r}\ln \left(\frac{r}{2UP}-1\right).
  \en
  It is clear that $v_g$ always less than 1 outside the horizon $r>2UP$. That is to say, the graviton propagates in the light cone along a timelike curve in the viewpoint of a flat spacetime.
  The physical interpretation is that the back reaction of this wave shrinks the light cone, so that the wave becomes null measured by the curved metric with full back reactions. A more tangible interpretation
  is $\frac{dr_*}{dr}>0$, thus the graviton travels less distance measured by $r$ than that of $r_*$  in the same time interval.  The effective mass of graviton reads,
  \be
  m_g^2=\omega^2-k^2=\omega^2(1-\frac{1}{v_g^2})<0.
  \en
 First, $m_g$ is different from the parameter $m$ in the action (\ref{poten1}). And it is variable with respect to all coordinates $u, r, \theta, \phi$.  Second, $m_g^2<0$ seems to implies an imaginary $m$. This phenomenon has drawn attentions in previous investigations of massive gravity \cite{imm}. Here we show that it is only
 an apparent difficulty rather an essential one, making no physical sense. The physical waves propagate along null geodesics in the full theory. Such an eccentric phenomenon appears only because we make a linear approximation. One sees again that linear approximation of a non-linear theory may leads to rather confusing results.

 The other difference between viewpoints of linear approximation and exact solution is the direction of the wave vector. In the viewpoint of exact solution, the wave vector reads,
  \be
  L^a=-g^{ab}(du)_b=\left(\frac{\partial}{\partial r}\right)^a.
  \en
  It denotes a wave propagating along $r$. However, in the viewpoint of exact solution, $r$ is no longer a radial coordinate since the spacetime may be deviated from spherical symmetry. In the viewpoint of linear approximation, it is a wave propagating on a Minkowski space in spherical coordinate system. Thus in this case $r$ is always the radial coordinate.

 To compared with other solutions of gravitational waves and black holes, we explore the algebraic property of this quasi-spherical wave spacetime. With this aim, we write the tetrad (\ref{tetrad1}) in Newman-Penrose formalism,
 \be
 g=-\tau_1\tau_2-\tau_2\tau_1+m\bar{m}+\bar{m}m,
  \label{tetrad2}
  \en
 where,
 \be
 m=\frac{1}{\sqrt{2}}(\tau_3-i\tau_4),~\bar{m}=\frac{1}{\sqrt{2}}(\tau_3+i\tau_4).
 \en

 The non-zero components of Weyl tensor read,
 \be
 \Psi_2=-C_{1342}=\frac{UP}{r^3},
 \en

 \be
 \Psi_3=-C_{1242}=\frac{3\sqrt{2}U}{4\sin\theta~ r^3}\left(i\frac{\partial P}{\partial \phi}-\sin \theta \frac{\partial P}{\partial \theta}\right),
 \en
 and,
 \be
 \Psi_1=\Psi_4=0.
 \en
 So it is a metric belonging to Petrov-II, just the same as topological spherical gravitational wave in $f(R)$-gravity \cite{self1}.  $\Psi_3$ vanishes when $P=$constant, i.e., the wave has no transverse polarization. In fact, it becomes a Vaidya metric, whose type is Petrov-D. The difference is that Vaidya metric is
 sourced by null dusts in GR. Here it is a vacuum space, in which a gravitational wave is propagating, in massive gravity.

 \section{Summary}

   We present a concise summary of this paper. Logically, massive gravity is one of the most natural extension of GR. In principle, one can obtain the mass of graviton through probing the velocity of gravitational waves. However, the definition of mass of graviton is illusive in massive gravity, which usually depends on coordinates \cite{imm}. The ambiguity roots in the fluctuations around a back ground spacetime, i.e., to decompose the full metric into a back ground and a wave. Further more, measurement of velocity of gravitational wave only can constrain the mass of graviton in an interval. It is extremely difficult to answer the question whether or not the mass of graviton is exactly zero through such a measurement. Aim to this problem, we have to find some non-perturbative property of massive gravity. We find such a property in this work, that is, a spherical/quasi-spherical gravitational wave is permitted in theory of massive gravity with a special singular reference metric. It is well-known that GR strictly prohibits spherical gravitational waves in vacuum space. Actually, it is not an easy task to find a sourced  spherical gravitational wave in GR.  With this solution, we clear some puzzles about the velocity and mass of gravitons. The imaginary mass problem of graviton is only an illusion yielded by linear approximations. The graviton rays are exactly null in the full metric and well-behaved. We demonstrate that massive graviton leads to spherical/quasi-spherical gravitational wave in vacuum space, no matter how tiny mass of the graviton is in dRGT theory with a singular reference metric. This may be treated as a fundamental discrepancy between massive gravity and GR.

 {\bf Acknowledgments.}
   H.Z. thanks Bin Zhou for helpful discussions. This work is supported in part by Shandong Province Natural Science Foundation under grant No.  ZR201709220395, and the National Key Research and Development Program of China (No. 2020YFC2201400).

\end{document}